\documentclass[12pt]{article}
\usepackage{epsfig,cite}
\usepackage {amsmath}
\usepackage{amssymb}
\usepackage{ulem}
\include{epsf}
\newcount\timecount
\newcount\hours \newcount\minutes  \newcount\temp \newcount\pmhours
\hours = \time \divide\hours by 60 \temp = \hours \multiply\temp by
60 \minutes = \time \advance\minutes by -\temp
\def\hour{\the\hours}
\def\minute{\ifnum\minutes<10 0\the\minutes
            \else\the\minutes\fi}
\def\clock{
\ifnum\hours=0 12:\minute\ AM \else\ifnum\hours<12 \hour:\minute\ AM
      \else\ifnum\hours=12 12:\minute\ PM
            \else\ifnum\hours>12
                 \pmhours=\hours
                 \advance\pmhours by -12
                 \the\pmhours:\minute\ PM
                 \fi
            \fi
      \fi
\fi }

\def\monthname{\relax\ifcase\month 0/\or January\or February\or
   March\or April\or May\or June\or July\or August\or September\or
   October\or November\or December\else\number\month/\fi}

\def\bold#1{\setbox0=\hbox{$#1$}%
     \kern-.025em\copy0\kern-\wd0
     \kern.05em\copy0\kern-\wd0
     \kern-.025em\raise.0433em\box0 }

\def\beq{\begin{equation}}
\def\eeq{\end{equation}}

\def\ga{\mathrel{\raise.3ex\hbox{$>$\kern-.75em\lower1ex\hbox{$\sim$}}}}
\def\la{\mathrel{\raise.3ex\hbox{$<$\kern-.75em\lower1ex\hbox{$\sim$}}}}
\def\gev{{\rm \, Ge\kern-0.125em V}}
\def\tev{{\rm \, Te\kern-0.125em V}}
\def\gyr{{\rm \, G\kern-0.125em yr}}

\def\gappeq{\mathrel{\rlap {\raise.5ex\hbox{$>$}}
{\lower.5ex\hbox{$\sim$}}}}
\def\lappeq{\mathrel{\rlap{\raise.5ex\hbox{$<$}}
{\lower.5ex\hbox{$\sim$}}}}
\def\Toprel#1\over#2{\mathrel{\mathop{#2}\limits^{#1}}}

\def\m12{m_{1\!/2}}
\def\bea{\begin{eqnarray}}
\def\eea{\end{eqnarray}}
\def\beq{\begin{equation}}
\def\eeq{\end{equation}}

\begin{document}
\begin{titlepage}
\pagestyle{empty} \baselineskip=21pt
%\rightline{\tt hep-ph/yymmnnn}
%\rightline{CERN-PH-TH/2011-???, KCL-PH-TH/2011-?} \vskip 0.2in
\rightline{KCL-PH-TH/2011-14, LCTS/2011-06, CERN-PH-TH/2011-131} \vskip 0.2in
\begin{center}
{\large{\bf Probing Lorentz Violation in Neutrino Propagation from a 
Core-Collapse Supernova}}
\end{center}
\begin{center}
\vskip 0.2in {\bf John~Ellis}$^{1,2}$, {\bf Hans-Thomas~Janka}$^{3}$,
{\bf Nikolaos~E.~Mavromatos}$^{1,2}$, {\bf Alexander~S.~Sakharov}$^{2,4}$ 
and {\bf
Edward~K.~G.~Sarkisyan}$^{2,5}$ \vskip 0.1in

{\it
$^1${Theoretical Particle Physics and Cosmology Group, Department of Physics, \\
King's College London, Strand, London WC2R 2LS, UK}\\
$^2${Physics Department, CERN, CH-1211 Geneva 23, 
Switzerland}\\
$^3${Max-Planck-Institut f\"ur Astrophysik, Karl-Schwarzschild-Str. 1, \\
D-85748 Garching, Germany}\\
$^4${Department of Physics, Wayne State University, Detroit, MI 48202,USA}\\
$^5${Department of Physics, University of Texas at Arlington, Arlington, 
TX 76019, USA}\\
}

\vskip 0.2in {\bf Abstract}
\end{center}
\baselineskip=18pt \noindent
%%%%%%%%%%%%%%%%%%%%%%%%%%%%%%%%%%%%%%%%%%%%%%%%%%%%%%%%%%%%%%%%%%%%%

Supernova explosions provide the most sensitive probes of
neutrino propagation, such as the possibility that neutrino
velocities might be affected by the foamy structure of space-time thought to be 
generated by quantum-gravitational (QG) effects. Recent two-dimensional 
simulations of the neutrino emissions from core-collapse supernovae suggest 
that they might exhibit variations in time on the scale of a few milliseconds. We analyze
simulations of such neutrino emissions using a wavelet technique, and consider the limits 
that might be set on a linear or quadratic violation of Lorentz invariance in the 
group velocities of neutrinos of different energies, $v/c = [1 \pm (E/M_{\nu {\rm LV}1})]$ or $[1 \pm (E/M_{\nu
{\rm LV}2})^2]$, if variations on such short time scales were to be 
observed, where the mass scales $M_{\nu {\rm LVi}}$ might appear in models of quantum gravity. We find
prospective sensitivities to $M_{\nu {\rm LV}1} \sim 2 \times 10^{13}$~GeV 
and 
$M_{\nu {\rm LV}2} \sim 10^{6}$~GeV at the 95~\% confidence level, up to
two orders of magnitude beyond estimates made using previous one-dimensional
simulations of core-collapse supernovae. We also analyze the prospective sensitivities to 
scenarios in which the propagation times of neutrinos of fixed energies are
subject to stochastic fluctuations.

\vspace*{0.2cm}

%%%%%%%%%%%%%%%%%%%%%%%%%%%%%%%%%%%%%%%%%%%%%%%%%%%%%%%%%%%%%%%%%%%%%%
%\vfill \leftline{CERN-PH-TH/2011-???} \leftline{May 2011}
\vfill \leftline{CERN-PH-TH/2011-131} \leftline{October 2011}
\end{titlepage}
%\baselineskip=18pt
%%%%%%%%%%%%%%%%%%%%%%%%%%%%%%%%%%%%%%%%%%%%%%%%%%%%%%%%%%%%%%%%%%%%%%

\section{Introduction}

\begin{figure}[htb]
\begin{center}
\includegraphics[width=0.8\textwidth]{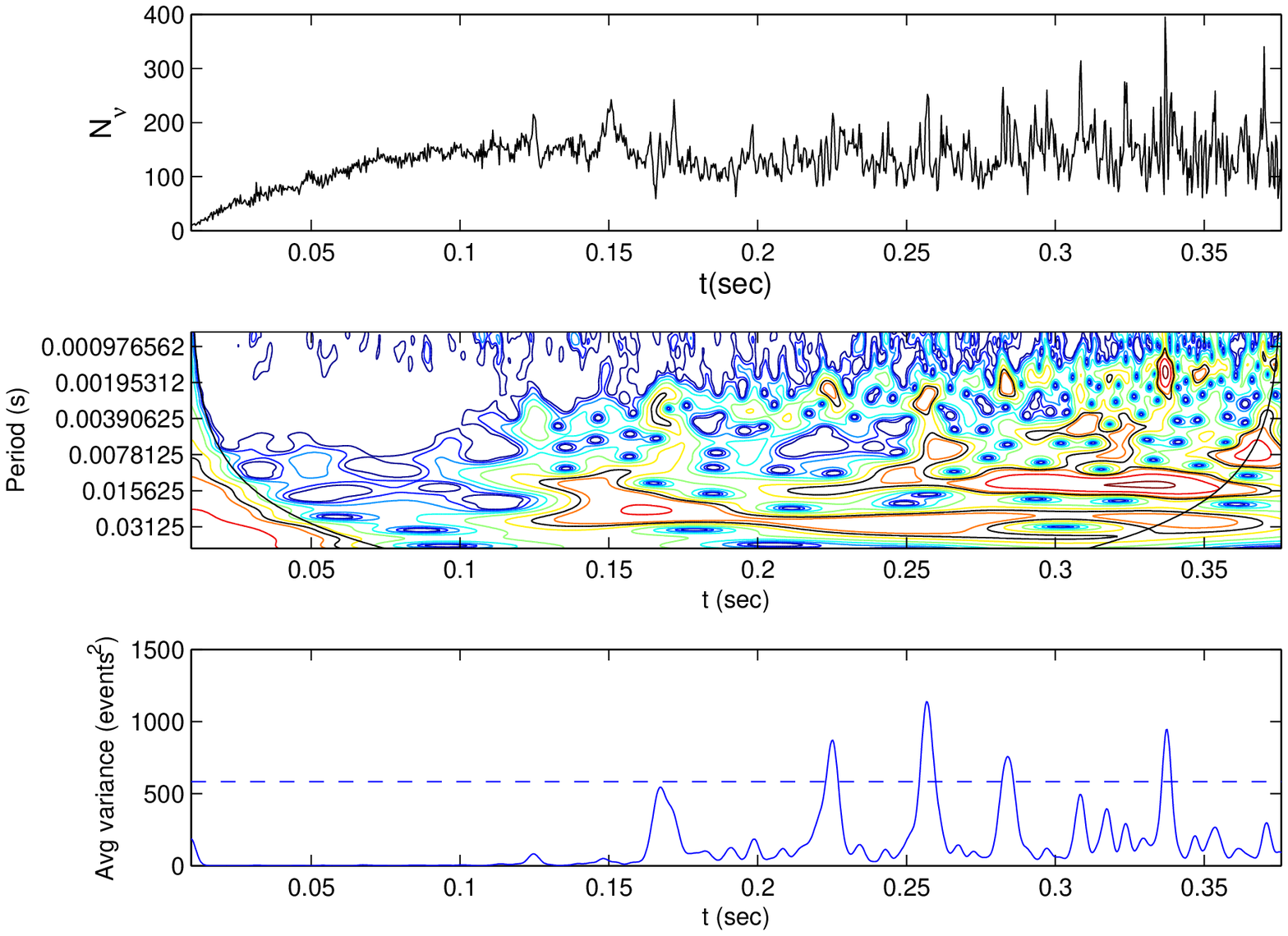}
\end{center}
\caption{\it 
{\underline {\it Top panel:}}
The time series of the neutrino emission from the
two-dimensional
simulation of
a core-collapse supernova found in~\protect\cite{2D}. The time
profile is sampled in 1024 ($2^{10}$) bins. 
{\underline {Middle panel:}} 
The local
wavelet power spectrum of the neutrino emission time series, obtained 
using the Morlet
wavelet function~(\protect\ref{morletWave}) normalized by $1/\sigma^2$.
The vertical axis is the
Fourier period (in seconds), and the horizontal axis is the time of the neutrino emission. 
The red contours enclose regions that differ from white noise at greater than the 95\% confidence
level. The cone of influence, where edge effects
become important, is indicated by the concave solid lines at the edges of the support of the
signal. Comparing the width of a peak in the
wavelet power spectrum with this decorrelation time,
one can distinguish between a spike in the data (possibly
due to random noise) and a harmonic component
at the equivalent Fourier frequency. 
{\underline {Bottom panel:}}
The average power in the 0.002 - 
0.003~s band. The dashed
line is the 95\% confidence level obtained from (\protect\ref{av3}).}
\label{fig:w_img}
\end{figure}

Supernovae provide some of the most sensitive
probes of neutrino physics~\cite{neutrinos}, as exemplified by studies of the
neutrinos detected after emission from SN 1987a~\cite{imbsn1987a}.
Excellent prospects for improving these studies would be
offered by a future galactic core-collapse supernova.
Crucial inputs into estimates of the prospective sensitivities
of such studies are provided by supernova simulations.
In the past, these simulations have mainly been one-dimensional, i.e.,
implicitly assuming that the collapse is spherically symmetric. Clearly,
more realistic simulations are desirable, and more recently two-dimensional
hydrodynamic models, i.e., implicitly assuming only cylindrical symmetry, 
have become available, in which multi-group, three-flavor neutrino transport
has been treated by different (relatively sophisticated) 
approximations~\cite{dataSim,dataSim2,2D}.
These have revealed several interesting features, the most
relevant for our analysis being the appearance of fast time variations
in the neutrino emission~\cite{2D,dataSim2}, 
as seen in the top panel of Fig.~\ref{fig:w_img}. 
It was shown that these could be observable in the IceCube
experiment~\cite{ice}, and could exhibit a quasi-periodicity of ${\cal O}(10)$~ms, 
similar to the natural time scale of reverberations associated with hydrodynamic
instabilities in the supernova core. It is
desirable to confirm the predicted appearance of such features in neutrino
emissions from core-collapse supernovae through more detailed simulations, 
in particular in three dimensions~\footnote{For first steps in this direction, see~\cite{3D-1,3D-2}.}.
Nevertheless, we find these features sufficiently interesting and well
motivated to consider the sensitivity to effects in neutrino 
propagation that would become available if such rapid time
variations were in fact to be observed.

The most obvious such effect is that of neutrino mass.
Non-zero masses cause neutrinos to travel at less than the speed
of light, by an amount that {\it decreases} with increasing neutrino
energy. This causes any time structure that appears simultaneously
in emissions over a range of energies to spread out before
arrival at the Earth, an effect that was exploited to set an upper
limit on neutrino masses using data from SN 1987a~\cite{numass1987}. Though an
interesting demonstration of principle, that limit
was not competitive with laboratory and cosmological limits,
and even the increase in sensitivity suggested by the more recent
two-dimensional simulations seems unlikely to be competitive.

Another possibility is to use the neutrino emissions from core-collapse
supernovae to constrain effects on neutrino propagation such as might be 
induced by `foamy' quantum-gravitational fluctuations in the fabric of space-time~\cite{foam}.  
Such space-time foam effects could include a refractive index  (i.e., a
change in the neutrino or photon velocity that depends on energy)~\cite{amellis,GRB,gambini,emnnewuncert,Harries}, dispersion in
propagation at fixed energy~\cite{mitsou}, and a loss of coherence~\cite{nudecoh}. Models suggest
that any such effects should {\it increase} with increasing neutrino energy $E$,
being proportional, e.g., to $E/M_{\nu {\rm LV}1}$ or $(E/M_{\nu 
{\rm LV}2})^2$, where the mass scales $M_{\nu {\rm LVi}}$ might originate from quantum gravity (QG). 
As such, they would
be easily distinguishable, in principle, from the effect on neutrino
propagation of a neutrino mass. Estimates have been given~\cite{Harries} of the
possible sensitivity to such effects that could be provided by a core-collapse
supernova explosion in our Galaxy, e.g., sensitivity to $M_{\nu {\rm LV}1} 
\sim 2 \times 10^{11}$~GeV or
$M_{\nu {\rm LV}2} \sim 2 \times 10^5$~GeV for a refractive index $> 1$, 
corresponding to
increasingly subluminal propagation of energetic neutrinos. These prospective
sensitivities are considerably greater than those offered by terrestrial long-baseline
neutrino experiments~\cite{MINOS}: the latters' beams have much finer time structures, but they are
handicapped by their much shorter propagation distances.

The estimates above were based on the earlier simulations of the core collapse of a supernova
that yielded emissions over a period of seconds without any finer time structure.
{\it A priori}, the observation of time structures on the scale of ${\cal O}(10)$~ms in the neutrino emissions
from core-collapse supernovae, as suggested by more recent two-dimensional 
simulations~\cite{2D,dataSim2},
would provide the possibility to constrain foamy effects on neutrino
propagation a couple of orders of magnitude more sensitively than was previously estimated.
In this paper, we use such a simulation to estimate the prospective sensitivity to a neutrino refractive index
and dispersion in propagation at fixed energy, applying a wavelet analysis to
the simulated neutrino signals published in~\cite{2D}.

We find that the prospective sensitivities to novel effects in neutrino propagation
would indeed be enhanced by two orders of magnitude if the neutrino signals
from core-collapse supernovae do exhibit the fine-scale time structures
suggested by two-dimensional simulations. In the first instance, these
sensitivities may be expressed for energy-dependent time shifts $\tau_n$
or dispersions at fixed energy, both expressible in units of s/MeV$^{\,l}$ 
for 
effects $\propto E^{\,l}$.
Knowing the distance from any given supernova, these sensitivities may be
translated into sensitivities to model parameters such as the $M_{\nu 
{\rm LV}l}$
introduced above. Hypothesizing a typical distance of 10~kpc we find, for
example, sensitivities to  $M_{\nu {\rm LV}1} \sim 2 \times 10^{13}$~GeV 
and 
$M_{\nu {\rm LV}2} \sim 10^{6}$~GeV at the 95~\% confidence level.
 We emphasize these sensitivities could immediately become {\it lower} limits if
time structures are observed. The absence of such time structures could provide
{\it upper} limits if the predictions by two-dimensional simulations of short time structures
could be validated, assuming that no astrophysical effect
during propagation could wash out the effect.
 % but that the absence of time structures could 
 %not be 
 %used as evidence for non-standard neutrino propagation unless
 %there is total theoretical confidence in the prediction of 
 %small time-scale structures in the collapse of the supernova 
 %core, e.g., on the  basis of other supernova observations or 
 %of three-dimensional simulations.

This subject has become much more topical during the finalization of this paper,
with the publication of an analysis of OPERA data~\cite{OPERA} using a technique similar
to that developed in~\cite{Harries}. This reports possible superluminal neutrino
propagation with velocity exceeding that of light, $c$, by an amount $\delta v$:
$\delta v/c \sim 2.5 \times 10^{-5}$, corresponding to $M_{\nu {\rm LV}1} \sim 1.1 \times 10^6$~GeV
or $M_{\nu {\rm LV}2} \sim 5.6 \times 10^3$~GeV~\cite{AEN}, smaller than the lower limits established
in~\cite{Harries} using SN1987a data. Thus, a simple power-law fit  jointly to the
SN1987a data and OPERA $\delta v$ would probably require an energy dependence steeper that $E^2$,
in apparent conflict with the energy spectrum of the neutrino events measured by OPERA.
The superluminal neutrino interpretation of the OPERA data is subject to many other
experimental and phenomenological constraints, and one should not assume that it will survive
further scrutiny. Nevertheless, this episode heightens awareness of the importance of probing
fundamental principles such as the universality of the velocity of light as sensitively as possible,
and our analysis based on two-dimensional simulations of supernova explosions shows that they could
provide unparallelled sensitivity to novel effects in neutrino propagation.

\section{Ingredients in the Analysis}

\subsection{Quantum-Gravity Models for Non-Standard Neutrino Propagation}
%\section{Quantum  Gravity Effects on Neutrinos}

As mentioned in the Introduction,
various possibilities for non-standard neutrino propagation are suggested by
phenomenological models based on approaches to quantum gravity~\cite{nudecoh,Harries}.
This is because such models entail microscopic fluctuations of space-time, due to
curvature fluctuations and/or, in certain theories, space-time defects. Specifically, in brane theories
based on string theory~\cite{emnnewuncert}, the latter may be modelled as point-like structures that cross
the brane Universe from the bulk, giving space-time a ``foamy'' nature at microscopic scales.
In our current state of knowledge of string theory, the string scale is essentially a free parameter 
to be constrained by low-energy phenomenology, and in particular by constraints on
non-standard neutrino propagation.

\subsubsection{Modified Dispersion Relation}

One of the most explored avenues for experimental probes of (some models of) space-time foam
is to search for Lorentz violation induced in the propagation of matter particles by their
interactions with this foamy space-time medium.
In certain string-inspired models of foam the presence of a medium affects the dispersion relations of certain
species of matter particles~\cite{mitsou,emnnewuncert}. In the simplest formulation of the effects of the foam, in a first approximation
only electrically-neutral particles interact with the medium, so that photons and neutrinos are
the most sensitive probes of such models. In these models, the modification of a particle's 
dispersion relation is a consequence of the microscopic Lorentz violation induced
by the recoil of a space-time defect during its non-trivial interaction with the open-string state that represents the 
particle excitation in a brane Universe~\cite{mitsou,emnnewuncert}. In the case of photon propagation,
this effect is manifested as a vacuum refractive index. For purely string-theoretical reasons, the induced
refractive index is {\it subluminal}, implying that, if a beam of photons 
with different energies is emitted 
simultaneously from a source, the arrival times of more energetic photons will be delayed compared to their lower-energy counterparts~\footnote{As a bonus, this avoids any constraints due to the emission of {\v C}erenkov radiation by energetic photons~\cite{cerenkov}.}.

In the case of neutrinos, regarded as (almost) massless particles, a similar effect
might be expected, namely a delayed arrival of the more energetic neutrinos from cosmic sources, assuming that the neutrinos of different energies are emitted 
(almost) simultaneously. One may therefore consider foam-induced Lorentz violation that is expressed via a
neutrino group velocity, $v_g$, that may depend either linearly or quadratically on the energy of the neutrino:
\begin{equation}
v_g/c = 1 \pm (E/M_{\nu {\rm LV}{\it l}})^l~, \quad l =1~{\rm  or}~ 2~.
\label{refr}
\end{equation}
As discussed in~\cite{Harries}, lower limits on $M_{\nu {\rm LV}l}$ may 
be 
obtained by requiring that narrow
peaks in neutrino emission over a range of energies not be broadened significantly, or even washed out,
and this is the strategy advocated here using emissions from a supernova.

\subsubsection{Dispersion of a Wave Packet}

A second effect that may also be induced by quantum-gravity foam models is  a spread in the width of the wave packet, which may
depend on the neutrino energy. This could arise from an energy-dependent neutrino velocity of the type discussed above, 
or from a stochastic spread in neutrino velocities at fixed energy~\cite{mitsou}.

Consider, for example, a neutrino wave packet that is Gaussian in the 
``approximately'' light-cone variable 
$x-v_g t$, where the group velocity $v_g = d\omega/d\kappa$ is near 
the speed of light for the relativistic, 
almost massless neutrinos that we consider here,
allowing for a generic dispersion relation
$\omega = \omega (\kappa)$ with $\kappa\equiv |\vec{\kappa}|$ being 
the 
spatial momentum 
amplitude. For neutrinos that are almost 
massless, the analysis is similar to that 
given in \cite{mitsou}. The square of the neutrino amplitude is given in general by:
\begin{equation}
   |f(x,t)|^2 = \frac{A^2}{
 \sqrt{1 + \frac{\alpha ^2 t^2}{(\Delta x_0)^4}}}\,
\exp\left\{-\frac{(x - v_g t)^2 }{2(\Delta x_0)^2
\left[1 + \frac{\alpha ^2 t^2}{(\Delta x_0)^4}\right]}\right\}~,
\label{timegauss}
\end{equation}
where 
$|\Delta x_0|$ is the spread at $t=0$,
$\alpha = \frac{1}{2} \left(d^2\omega /d^2 \kappa\right)$, and 
 we assume that the neutrino wave packet has a Gaussian distribution in
the ``approximately'' light-cone variable $x-v_g t$.

We see immediately in (\ref{timegauss}) that the quadratic term
$\alpha$ in the dispersion relation
does not affect the motion of the peak, but only the spread of the
Gaussian wave packet:
\begin{equation}
 | \Delta x | =
\Delta x_0 \sqrt{1 + \frac{\alpha ^2 t^2}{(\Delta x_0)^4}}~,
\label{spread}
\end{equation}
which therefore increases with time. The quadratic term $\alpha$ also affects
the peak amplitude of the wave packet:
the latter decreases as the spread
(\ref{spread}) increases, in such a way that the integral of $|f(x,t)|^2$ is constant.

If the neutrinos were exactly massless, then, as in the case of photons, the quantity $\alpha$ would receive 
non-zero contributions only from anomalous terms in the dispersion relation due to quantum gravity of the form 
(\ref{refr})~\cite{mitsou}. In the linear case, $l=1$, such corrections 
would be independent of the energy of the neutrino.
In the presence of small neutrino masses, $m \ll \kappa$, there are always 
contributions to $\alpha $ from terms of the 
form:
\begin{equation}
\alpha = \frac{m^2}{\kappa^3}~, \quad m \ll \kappa~.
\end{equation}
Such terms contribute to the spread of the wave packet, but decrease with 
the neutrino momentum, in contrast to quantum-gravity effects that are 
expected to be constant or to increase as functions of
the neutrino energy (momentum), depending whether the energy dependence of the refractive index
is linear or quadratic.

Hence we may parametrize the spread $\alpha$ generically as in 
(\ref{spread}), with the parameter $\alpha$
having a power-law dependence on the neutrino momentum $\kappa$:
\begin{equation}
\label{alphavalue}
\alpha = \frac{m^2}{\kappa^3} - l\, (l + 1) \frac{\kappa^{\, l-1}}{M_{\nu 
{{\rm LV}{\it l}}}^l}~,
\end{equation}
with $l=1$ or $2$, where the stochastic case is denoted by the tilde 
in the suffix of the QG scale. To leading 
order in the small neutrino mass, we may replace $k$ by the (average) 
neutrino energy $E$ of the wave packet. This effect leads to a potentially independent  way of detecting 
space-time foam~\cite{foam}, although we find that it is not competitive with limits coming
directly from time-of-flight measurements, as discussed below.

\subsubsection{Wave-Packet Spread induced by Stochastic Fluctuations in Neutrino Velocities}

Another phenomenon that may also be induced by quantum-gravity foam models is stochastic fluctuation in the
velocities of different neutrinos with the same energy~\cite{mitsou}. As an example how this type of effect might
arise  from space-time foam, we consider the possibility of light-cone fluctuations. 

In the string-inspired models of space-time foam that we consider here, these may be induced by
the summation over world-sheet surfaces with higher-genus topologies. These result in an
effective stochastic fluctuation of the light cone of order~\cite{mitsou}
\begin{equation}\label{stochc}
\delta c \sim 8 g_s^2 \frac{E}{M_s c^2}~,
\end{equation}
where $M_s$ is the string scale, $g_s$ is the string coupling and $E$ 
is the average energy of the massless (or, in the case of the neutrino,
almost massless) probe.

Light-cone fluctuations of the form (\ref{stochc}) would also lead to a spread in the Gaussian wave packet, 
which is distinct from the spread induced by the refractive index (\ref{spread}). The spread induced by
light-cone fluctuations would be linear in the quantum gravity scale, as seen from (\ref{stochc}). 
Such an effect would lead to a stochastic spread in the arrival times of photons or neutrinos of order
\begin{equation}\label{stochc2}
\delta \Delta t = \frac{L}{c\Lambda} E~, \quad \Lambda \equiv \frac{M_s c^2}{8g_s^2}~,
\end{equation}
where $L$ is the distance of the observer from the source~\footnote{In our analysis below we may ignore effects 
associated with the expansion of the Universe, as we are dealing with neutrinos from galactic supernovae.}.
We emphasize that, in contrast to the variation (\ref{refr}) in the refractive index -
which refers to photons of different energies - the fluctuation (\ref{stochc2})
characterizes the statistical spread in the velocities of particles {\it of the same energy}.
We note that the stochastic effect (\ref{stochc2}) is suppressed compared to the linear ($n=1$) refractive index
effect (\ref{refr}) by an extra power of the string coupling $g_s$ (we recall that,
in the string model, $M_{\nu {\rm LV}1} \propto M_s/g_s$).

The light-cone fluctuation effects may be thought of as inducing a 
time-independent spread $\sigma$ in a neutrino 
wave-packet that can be parametrized as (see later)
\begin{equation}\label{timeprofile}
\mathcal{P}(t) \sim   e^{- \frac{(t - t_0)^2}{2\sigma^2}}~,
\end{equation}
where
 %In this sense, the light cone fluctuation effects add to a constant 
 %$\sigma^2$ foam-induced terms proportional to $E$ or $E^2$ (in the 
 %string model, discussed explicitly above is just linear):
\begin{equation}\label{spread2}
\sigma^2 = \sigma_0^2 + c_1^2 \frac{E^{\,l}}{M_{\nu \, 
\widetilde{{\rm LV}l}}^{\,l}}~, \quad l =0~{\rm or}~1~.
\end{equation}
The expression (\ref{timeprofile}) is distinct from the relativistic Gaussian wave packet 
 %I 
used above and 
 %used 
in \cite{mitsou}. In the latter case, the stochastic light cone 
fluctuations already affect the spread (\ref{spread}) $|\Delta x_0|^2$ at 
$t=0$, which then may be identified with the $\sigma^2$ in (\ref{spread2}).

\subsection{Two-Dimensional Simulation of a Core-Collapse Supernova \label{SASI}}

The analysis presented in this paper is based on the results of a
two-dimensional, i.e., axi-symmetric,
core-collapse simulation for a 15\,$M_\odot$ star computed with the 
high-density equation of state of Lattimer \& Swesty~\cite{lattimer}.
The properties of the neutrino signal calculated in this model were
discussed in detail in~\cite{2D}. 

Unlike one-dimensional, i.e., spherically-symmetric, models, the neutrino
emission during the post-bounce accretion phase exhibits rapid 
time-variability because of anisotropic mass flows in the accretion
layer around the newly-formed neutron star. These flows
are a consequence of convective overturn as well as the
standing accretion-shock instability (SASI; \cite{SASI}), which 
lead to large-scale, non-radial mass motions in the
layer between the proto-neutron star surface and the accretion shock. 
Locally-enhanced mass infall to the compact remnant and asymmetric 
compression create hot spots that can produce transiently 
neutrino radiation that is more luminous and with a harder spectrum, emitted in preferred
directions. Temporal variations of the luminosities and mean energies
are expected to persist during the whole accretion phase,
which can last hundreds of milliseconds. For electron neutrinos 
and antineutrinos, such variations could yield fractional changes 
of 10\% and even higher during the most violent phases of core activity
in two-dimensional models with no or only slow rotation~\cite{2D,dataSim2}.
The corresponding effects are somewhat
damped for muon and tau neutrinos, because a smaller fraction of these neutrinos is produced in the
outer layers of the proto-neutron star where asymmetric accretion
causes the largest perturbations.

The fluctuating neutrino emission has been shown to trigger 
a clearly detectable signature in the response of the IceCube
detector in the case of a neutrino burst from a future
Galactic supernova. Peaks of the power spectrum of the
event rate are expected at the typical frequencies of
the SASI and convective activity (between several tens of Hz and 
roughly 200~Hz)~\cite{2D}.

While a softer nuclear equation of state (allowing for a more
compact proto-neutron star) seems to favor larger signal
amplitudes~\cite{2D}, several other sources of uncertainty 
in the model predictions need to be mentioned.
One concerns the neutrino transport description that is used,
which even in the most sophisticated current multi-dimensional 
(two- or three-dimensional) models cannot be handled without approximations.
In Refs.~\cite{2D}, neutrino predictions were obtained in a scheme
in which the full energy dependence of the transport problem
was accounted for (including D\"oppler shifting, gravitational
redshifting, and neutrino redistribution in energy space by
scattering reactions) but the two-dimensionality of the 
transport was treated in a ``ray-by-ray'' approximation.
This means that the spatial transport was
described by $N$ spherically symmetric (radial) problems with
$N$ being the number of lateral grid zones, and it implies that
the directional averaging or smoothing of the neutrino
emission due to radiation received by an observer from 
different areas of the emitting surface is underestimated.
However, in Refs.~\cite{2D} also data averaged over hemispheres 
(northern, southern, equatorial) were considered, which still led 
to easily-detectable signatures. Moreover, the basic effects discussed
in~\cite{2D} were confirmed by true multi-angle 2D transport results
in Refs.~\cite{dataSim2}, though with the limitation of not including energy-bin
coupling due to the effects mentioned above. We note also that flavour oscillations between 
$\bar\nu_e$ and muon or tau anti-neutrinos, whose emission
variations have lower amplitudes, reduces the modulations of
the event rate in the IceCube detector only moderately.

The largest uncertainty in current model predictions 
results from the two-dimensional nature of the most detailed 
simulations. Three-dimensional modelling of stellar core
collapse is still in its infancy, and well-resolved simulations with energy-dependent neutrino transport
are currently not available. First steps in this direction were 
reported in Refs.~\cite{3D-1}, but the models are either
not evolved for interestingly long post-bounce times or
the employed numerical resolution is poor and the neutrino
data are not conclusive with respect to the effects discussed here.
In constrast, Ref.~\cite{3D-2} employed a simpler, energy-averaged
(``grey''), ray-by-ray neutrino transport approximation and could
follow the evolution of collapsing stellar cores in 3D over several 
hundred milliseconds of post-bounce accretion, through explosion, 
into the subsequent neutrino-cooling of the nascent neutron star.
The radiated neutrino signal as visible by a distant observer was 
evaluated in~\cite{3D-2} and also revealed variations with time
on the scale of a few milliseconds, however with an amplitude
of several percent only, instead of the 10\% or more found in two-dimensional simulations. 

An analysis
of the detectable consequences is in progress. Whilst in a two-dimensional simulation
the existence of a symmetry axis directs the SASI sloshing 
motions of the shock and of the accretion flows, these motions
are similar in all directions in three dimensions and thus appear to develop
smaller amplitude in any particular direction, leading to a 
reduced fractional fluctuation of the observable neutrino
emission. It should be noted, however, that the existing
three-dimensional models still contain severe approximations and do not
explore the range of interesting possibilities. In particular,
they do not include the effects of rotation in the stellar core,
which even for slow rates could significantly influence the
growth of SASI spiral modes~\cite{SASI-spiral}.

We base our analysis here on the more mature two-dimensional simulations,
though adopting a somewhat optimistic point of view, in that we determine
the maximal effects that can be expected within the detailed two-dimensional models currently
available. We therefore consider (north-)polar
emission (i.e., no averaging of luminosities and spectra over a wider
range of latitudes) of electron antineutrinos, as
predicted by the 15\,$M_\odot$ simulation with the relatively
soft equation of state of Lattimer and Swesty in Refs.~\cite{2D}.
Possible flavor conversions between electron antineutrinos and
heavy-lepton antineutrinos are ignored.

\subsection{Detector response to the time-varying SN neutrino signal}

As was discussed in detail in~\cite{2D}, IceCube or a future megaton-class water
{\v C}erenkov detector would be very promising for detecting the time-varying neutrino signal from a 
future galactic SN. Such detectors are designed to detect a large number of {\v C}erenkov
photons produced by neutrino events, and a single photon produced by a given neutrino can 
tag its arrival time. Hence the term `event' can be used interchangeably to refer to photon
or neutrino detection. In the case of a SN at the
fiducial distance of $10$~kpc assumed here, the photon
detection rate can be as high as $\sim 10^3$/ms. This is similar to the intrinsic
background rate estimated for IceCube. A megaton-scale water {\v C}erenkov detector would
achieve neutrino detection rates similar to IceCube and, in addition, would provide event-by-event
information. Therefore, IceCube may serve as a benchmark detector for estimating a typical 
detection rate achievable for a time-varying SN neutrino signal~\footnote{For the Super-Kamiokande detector with fiducial volume
22.5~kt, the corresponding neutrino detection rate is approximately 2 orders of magnitude smaller,
but essentially background free.}.    

A schematic model of the IceCube detector response to a SN neutrino
signal was used in~\cite{2D} to estimate the detection rate, including efficiencies, for
{\v C}erenkov photons originating from the dominant inverse-beta reaction $\bar\nu_e +p\rightarrow n+e^+$:
\beq
\label{detRate1}
R_{\bar\nu_e}=114\ {\rm ms}^{-1}\frac{L_{\bar\nu_e}}{10^{52}\ {\rm erg\ s^{-1}}}
\left(\frac{10\ {\rm kpc}}{D}\right)^{2}\left(\frac{E_{\rm rms}}{15\ {\rm MeV}}\right)^{2},
\eeq
where $L_{\bar\nu_e}$ and $D$ are the SN luminosity and distance, respectively. The 
definition~\cite{2D} 
\beq\label{detRate2}
E_{\rm rms}^2 \equiv \frac{<E^3>}{<E>}
\eeq 
is used, where the average is to be taken over the neutrino distribution function. This estimate for the
photon count rate uses an approximate inverse beta cross section of 
$\sigma = 9.52\cdot 10^{-44}\ {\rm cm^2}(E_{\bar\nu_e}/{\rm MeV}^2)$.  

\begin{figure}[htb]
\begin{center}
\includegraphics[width=0.8\textwidth]{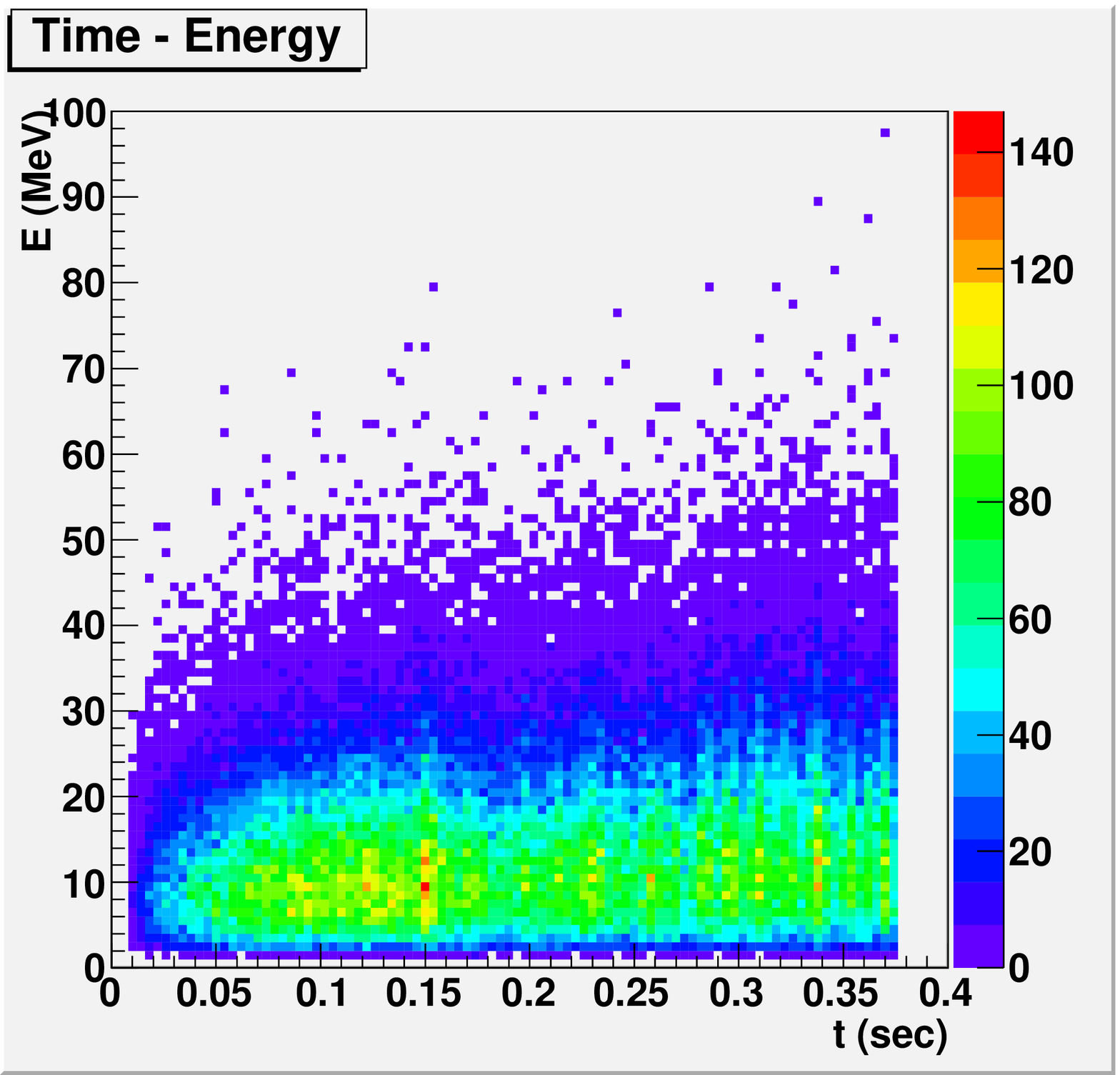}
\end{center}
\caption{\it The distribution of times and energies assigned to individual neutrinos
in one statistical realization of the thermal spectra found in the simulation~\protect\cite{2D}.}
\label{fig:ET}
\end{figure}

We assume in our analysis that the neutrino data collected from a supernova
explosion will consist of a list of individual neutrino events with
measured energies $E_i$ and arrival times $t_i$, as motivated by the fact that a
low-energy water {\v C}erenkov or scintillator detector is able to register the time $t_i$ of every event
with high precision.  
The results of the simulation performed along the line of subsection \ref{SASI} and described
extensively in~\cite{2D} are
presented as a set of primary energy fluxes within time periods of durations $\simeq 3 - 5~{\rm ms}$,
and each individual flux
can be represented by a black-body spectrum with a given value of the mean energy. The fluxes are
mapped into the photon counting rates using the benchmark detector response rate (\ref{detRate1}).
Knowing the mean and total energy of neutrinos leading to the photon counting rate in each time period, we
assign statistically to each event a specific time of emission and energy. 
The distribution of one implementation of such neutrino time-energy assignments folded with the 
benchmark detector response (\ref{detRate1}) is presented in Fig.~\ref{fig:ET}. 
In order to obtain robust estimates of the sensitivities to novel effects in neutrino propagation as
discussed below, we make a number of different implementations of the neutrino emission, all with
independent statistical realizations of the thermal spectra. We then apply a technique based on
wavelet transforms to these implementations, and analyze statistically the sensitivities to
new physics that they may have.

\subsection{Wavelet Analysis Technique}

We use a wavelet transform technique (see~\cite{malat} for a review) to analyze the neutrino time series
generated by the simulated supernova
explosion, as it is well adapted to capturing possible signatures of non-stationary
power at many different frequencies. 
 
Consider a time series, $x_n$, $n=0,\dots , N-1$, 
in bins of equal width $\delta t$.
 The wavelet transform is 
based on a wavelet
function, $\psi_0 (\eta )$, that depends on a dimensionless ``time"
parameter $\eta$. To be admissible as a wavelet transform, this function must have zero
mean and be localized in both time and frequency space. 
In choosing the wavelet function, there are several factors which should be
considered:

{\it Non-orthogonality}: The term ``wavelet function"
may be applied generically to
either orthogonal or non-orthogonal wavelets. In orthogonal
wavelet analysis, the number of convolutions at
each scale is proportional to the width of the wavelet
basis at that scale. This produces a wavelet spectrum
that contains discrete blocks of wavelet
power, and is useful for signal processing as it gives
the most compact representation of the signal. Unfortunately
for time series analysis, an aperiodic
shift in the time series produces a different wavelet
spectrum. Conversely, a non-orthogonal analysis
is highly redundant
at large scales, where the wavelet spectrum at adjacent
times is highly correlated. The term ``wavelet basis" refers
only to an orthogonal set of functions, and an orthogonal basis implies
 the use of the discrete wavelet transform, whereas non-orthogonal wavelet
functions can be used with either discrete or continuous wavelet
transforms.  A non-orthogonal
transform is useful for the analysis of time series
where smooth, continuous variations in wavelet
amplitude are expected, and is used in this study.

{\it Complexity}: A real wavelet function provides only a
single component, and can be used to isolate peaks
or discontinuities. On the other hand, a complex wavelet function
provides information about both the amplitude and phase, and is better adapted for
capturing oscillatory behavior. This is the choice made in this paper.

{\it Width}: For concreteness, the width of a wavelet
function is defined as the $e$-folding ``time" of the
wavelet amplitude. The resolution of a wavelet
function is determined by the balance between the
width in real space and the width in Fourier space.
A narrow (in ``time") function will have good time
resolution but poor frequency resolution, while a
broad function will have poor time resolution, but
good frequency resolution.

{\it Shape}: The wavelet function should reflect the type
of feature present in the time series. For time series
with sharp jumps or steps, one would choose
a boxcar-like function, whereas for a
smoothly-varying time series one would choose a
smooth function such as a damped cosine. If one
is primarily interested in wavelet power spectra,
then the choice of wavelet function is not critical,
and one function will give qualitatively similar
results to another.

Among common non-orthogonal wavelet functions, the Morlet wavelet
is complex and contains a number of oscillations sufficient to detect 
narrow features of the power spectrum, and is the choice made here.
It consists of a plane wave modulated by a Gaussian function in a variable $\eta$:
\begin{equation}\label{morletWave}
\psi_0(\eta )=\pi^{-1/4}e^{i\omega_0\eta}e^{-\eta^2/2},
\end{equation} 
where $\omega_0$ is a dimensionless frequency.

The continuous wavelet transform of a discrete sequence $x_n$ is defined as the
convolution of $x_n$ with a scaled and translated version of 
$\psi_0(\eta )$:~\footnote{The subscript 0 on $\psi$ has been dropped, in 
order to indicate that $\psi$ has also been normalized (see later).}
\begin{equation}\label{transform}
W_n(s)=\Sigma_{n'=0}^{N-1}x_{n'}\psi^*\left[\frac{(n'-n)\delta t}{s}\right] .
\end{equation}
By varying the wavelet scale $s$ and translating along the localized time 
index $n$, one can construct a picture showing both the amplitude of any
features versus the scale and how this amplitude varies with
time. Although it is possible
to calculate the wavelet transform using (\ref{transform}), it is convenient
and faster to perform the calculations in Fourier space.

To approximate the continuous wavelet transform, the convolution
(\ref{transform}) should be performed $N$ times for each scale, where $N$ is the 
number of points in the time series. By choosing $N$ points, the convolution
theorem allows us to perform all $N$ convolutions simultaneously in Fourier space
using a discrete Fourier transform (DFT):
\begin{equation}\label{dft1}
\hat x_k=\frac{1}{N}\Sigma_{n=0}^{N-1}x_{n}e^{-2\pi ikn/N},
\end{equation}  
where $k=0,\dots, N-1$ is the frequency index. In the continuous limit, 
the
Fourier transform of a function $\psi (t/s)$ is given by $\hat\psi (s\omega )$.
According to the convolution theorem, the wavelet transform is the Fourier
transform of the product:
\beq\label{dft2}
W_n(s)=\Sigma_{k=0}^{N-1}\hat x_k\psi^*(s\omega_k)e^{i\omega_kn\delta t},
\eeq
where $\omega_k=+\frac{2\pi k}{N\delta t}$ 
 and $-\frac{2\pi k}{N\delta t}$
for $k\le\frac{N}{2}$ and
$k>\frac{N}{2}$, respectively.

As already mentioned, following the
criteria for selecting wavelets for a particular task described above, 
in this paper we process the neutrino signal using Morlet wavelets 
(\ref{morletWave})
 of frequency $\omega_0$,
 %:
 %\beq\label{morl1}
 %\psi_0(\eta )=\pi^{-1/4}e^{i\omega_0\eta}e^{-\eta^2/2},
 %\eeq  
 which takes the form
\beq\label{morl2}
\hat\psi_0(s\omega )=\pi^{-1/4}H(\omega )e^{-(s\omega -\omega_0)^2/2}
\eeq
after the Fourier transform, 
where $H(\omega )$ is the Heaviside function: $H(\omega)=1 $ if $\omega > 0$, and zero otherwise. 
The width of this wavelet, defined as the 
$e$-folding time of the wavelet amplitude, is $\tau_s=\sqrt{2}s$.  
The function $\hat\psi_0$ is normalized to unity:
\beq\label{norm2}
\int|\hat\psi_0(\omega')|^2d\omega'=1
\eeq
and, in order to ensure that the wavelet transforms (\ref{dft2}) at all scales $s$ are
directly comparable to each other and to the transforms of other time series,
the wavelet functions $\psi_0$ at other scales are normalized to have unit energy:
\beq\label{norm1}
\hat\psi (s\omega_k)=\sqrt{\left(\frac{2\pi s}{\delta t}\right)}
\hat\psi_0(s\omega_k) .
\eeq
This implies that
\beq\label{norm3}
\Sigma_{k=0}^{N-1}|\hat\psi (s\omega_k)|=N. %,
\eeq
 %where $N$ the number of points. 
Using the convolution formula
(\ref{transform}), the normalization of the function $\psi$ is
\beq\label{norm4}
\psi\left[\frac{(n'-n)\delta t}{s}\right]=\sqrt{\frac{\delta t}{s}}\:\:
\psi_0\left[\frac{(n'-n)\, \delta t}{s}\right] ,
\eeq
and the wavelet power spectrum is defined by $|W_n(s)|^2$. It is desirable to
find a common normalization for the wavelet spectrum. Using the normalization
in (\ref{norm1}), and refering to (\ref{dft2}), the expectation value for
$|W_n(s)|^2$ is equal to $N$ times the expectation value for $|\hat x_k|^2$. For
a white-noise time series, this expectation value is $\sigma^2/N$, where
$\sigma^2$ is the variance. Thus, for a white-noise process, the expectation
value for the wavelet transform is $|W_n(s)|^2=\sigma^2$ for all $n$ and 
$s$. 

Once the wavelet function is chosen, it is necessary 
to choose a set of scales $s$ 
to use in (\ref{dft2}). For our purposes, it is convenient to 
choose discrete scales related by powers of two:
\beq\label{scales1}
s_j=2^{j\: \delta j}s_0,\qquad j=0,1,\dots ,J, 
 %%\eeq  
 %%\beq\label{scales2}
\qquad J={1\over \delta j}\,\log_2\left(\frac{N\delta t}{s_0}\right),
\eeq
where $s_0$
 %~\footnote{$s_0$ should be choosen so that the equivalent 
 %Fourier period,
 %which is $\lambda =\frac{4\pi s}{\omega_0+\sqrt{2+\omega_0^2}}$ 
 %for Morlet wavelet in use, is
  %approximately $2\delta t$.} 
 is
the smallest resolvable scale and $J$ determines the largest 
scale. The choice of a sufficiently small $\delta j$ depends on the width in spectral-space of the
wavelet function. In the case of the Morlet wavelet, $\delta j\approx 0.5$ is the 
largest value that still gives
 an adequate sampling scale.  In the middle panel of Fig.~\ref{fig:w_img} we 
use: 
$N=1024$, 
$\delta t=1.785\cdot 10^{-4}$~s, $s_0=2\delta t$, $\delta j=0.125$ and $J=48$.
We display in this panel the ``cone of influence", which
is indicated by the concave solid lines at the edges of the support of the
signal: this is the region of the wavelet
spectrum where edge effects
become important, defined as the $e$-folding
time for the wavelet autocorrelation power at
each scale, and also
gives a measure
of the decorrelation time for a single spike in the
time series.

Since the wavelet transform is a bandpass filter with
a known response function (the wavelet function), it
is possible to reconstruct the original time series using
either deconvolution or the inverse filter. In the case 
considered here, the reconstructed time series  
 %is just 
 can be represented as
 the sum
of the real parts of the wavelet transforms over all scales~\cite{Farge1992}:
\beq\label{rec1}
x_n=\frac{\delta j \, \sqrt{\delta t}}{C\,\psi_0(0)}
\Sigma_{j=0}^{J}\frac{{\rm Re}[W_n(s_j)]}{\sqrt{s_j}} .
\eeq
The factor $\psi_0(0)$ removes the energy scaling while the 
$\sqrt{s_j}$
 converts the wavelet transform to an energy density.
The factor $C$ comes from the reconstruction of a
$\delta$ function from its wavelet transform using the function 
$\psi_0(\eta )$ (\ref{morletWave}) 
 %%. This $C$ 
and 
 is a constant for each type 
of wavelet function.
The total energy is conserved by the wavelet transform, and the
equivalent of Parseval's theorem for wavelet analyses is
\beq
\label{recParseval}
\sigma^2=\frac{\delta j\delta t}{C\, N}\Sigma_{n=0}^{N-1}
\Sigma_{j=0}^{J}\frac{|W_n(s_j)|^2}{s_j}.
\eeq
 %%where $\sigma^2$ is the variance and $\delta$ function is assumed for 
 %%reconstruction. 

To determine significance levels for either Fourier
or wavelet spectra, one first needs to choose an appropriate
background spectrum. It is then assumed that
different realizations of the neutrino emission process will
be randomly distributed about this mean or expected
background, and the actual spectrum is compared
with this random distribution. For our phenomena, an appropriate background spectrum could be either
white noise (with a flat Fourier spectrum) or
red noise (increasing power with decreasing frequency). Here, for 
simplicity we choose 
a Gaussian white-noise background spectrum. 

We define as follows the null hypothesis for the wavelet power spectrum: we 
assume that the time series
has a mean power spectrum, 
given simply by $P_k=1$ in case of the white noise. If a peak in the wavelet
power spectrum appears significantly above this background spectrum, then it 
 is %%can be assumed 
 considered 
 to be a true
feature with a certain percentage confidence. If $x_n$ is a normally-distributed random variable, then
both the real and imaginary part of $\hat x_k$ are normally distributed. 
Since the 
square of a normally distributed variable is chi-squared distributed
with one degree of freedom (DOF), then the $|\hat 
x_k|^2$ variable has a 
chi-squared 
distribution with two DOFs, 
 %denoted by 
 $\chi_2$~\cite{Jenkins}. In the case that the original Fourier components are 
normally 
distributed, the
wavelet coefficients should also be normally distributed, while the 
wavelet power
spectrum $|W_n(s)|^2$ should have a $\chi_2^2$ distribution. Thus, 
if the background were truly white-noise, 
the distribution
shown in the middle panel of Fig.~\ref{fig:w_img} would have a
 $\chi_2^2$ distribution 
for each 
point of $(t,s)$. 
In summary, assuming a mean background spectrum of 
white (red) noise form,
the distribution of the Fourier power spectrum reads: 
\beq\label{cl1}
\frac{N|\hat x_k|^2}{\sigma^2}\Rightarrow P_k\chi_2^2 
\eeq 
at each Fourier frequency index $k$, with $P_k$ being  
the
mean spectrum value corresponding to the wavelet scale $s$ at this index. 
(Here the 
sign $\Rightarrow$ means
``is distributed as''.) The corresponding
distribution for the local wavelet power spectrum is:
\beq\label{cl2}
\frac{|W_n(s)|^2}{\sigma^2}\Rightarrow\frac{1}{2}P_k\chi_2^2
\eeq
at each time $n$ and scale $s$. 
 Disregarding the relation between $k$ and $s$, the relation (\ref{cl2}) is 
independent
of the wavelet function. After finding an appropriate background spectrum,
simply 
the white nose
in our case, and choosing a particular confidence level for
$\chi^2$ such as 95\%, one can then use (\ref{cl2}) at each scale
and build 95\% contour lines, as seen in the middle panel of Fig.~\ref{fig:w_img}.

In order to examine fluctuations in power over a range of
scales $(s_1, s_2)$ (a band), one can define the scale-averaged
wavelet power as the weighted sum of the wavelet
power spectrum over the scales $s_1$ to $s_2$:
\beq\label{av1}
\bar W_n^2=\frac{\delta j\, \delta t}{C\,}
\Sigma_{j=j_1}^{j_2}\frac{|W_n(s_j)|^2}{s_j} .
\eeq
The scale-averaged wavelet power can
be used to examine modulation of one time series by
another, or modulation of one frequency by another
within the same time series.

It is convenient
to normalize the wavelet power by the expectation
value for a white-noise time series. From (\ref{av1}), this
expectation value is $(\delta j\: \delta t\, \sigma^2)/(C\, S_{\rm
avg})$,
   %%~\footnote{The factor $C$ is a constant for each wavelet 
   %function
  %$\psi_0(\eta )$, and comes from the reconstruction of a
  %$\delta$ function from its wavelet transform using the function 
 %$\psi_0(\eta )$.}
where
 %$\sigma^2$ is the time-series variance and 
$S_{\rm avg}$ is defined as 
\beq\label{av2}
S_{\rm avg}=\left(\Sigma_{j=j_1}^{j_2}\frac{1}{s_j}\right)^{-1} .
\eeq
Using the normalization factor for white noise, the
distribution can be modelled in analogy with (\ref{cl2}), namely 
\beq\label{av3}
\frac{C\, S_{\rm avg}}{\delta j\, \delta t\, \sigma^2}\bar W_n^2 
\Rightarrow 
\bar
P\, \frac{\chi_{\vartheta}^2}{\vartheta} ,
\eeq
where the scale-averaged theoretical spectrum is now
given by
\beq\label{av4}
\bar P=S_{\rm avg}\Sigma_{j=j_1}^{j_2}\frac{P_j}{s_j} .
\eeq
and $\chi_{\vartheta}^2$ is the chi-squared distribution with 
 the number of DOFs $\vartheta$.
  %%The value of $P_j$ corresponds to the mean spectrum at the 
  %%Fourier frequency $j$
  %that corresponds to the 
  %%of the wavelet scale $s_j$.
 Note that for white noise the normalization is such that this spectrum is still unity. 
 The number of DOFs %degrees of freedom 
 $\vartheta$
in (\ref{av3}) is modelled as
\beq\label{av5}
\vartheta =\frac{2n_{\rm a}\,S_{\rm avg}}{S_{\rm mid}}\sqrt{1+
\left(\frac{n_{\rm a}\delta j}{\delta j_0}\right)^2},
\eeq
where $S_{\rm mid}=s_02^{0.5(j_1+j_2)\delta j}$ 
 and $n_{\rm a}=j_2-j_1+1$. 
 The factor $S_{\rm 
avg}/S_{\rm
mid}$ corrects for
the loss of 
 DOFs   
 %degrees of freedom 
 that arises from dividing the wavelet power spectrum by the scale in 
(\ref{av1}).

\section{Results of the Analysis}

\subsection{Wavelet transforms of the neutrino time series}

The neutrino time series found in~\cite{2D}, summing over all the produced neutrino energies, is shown
in the top panel of Fig.~\ref{fig:w_img}~\footnote{The binning is choosen at the level of
a good fraction of ms, which seems to be reasonable for water {\v C}erenkov and
scintillator detectors.}.
We see that it exhibits structures on time scales below a hundredth of a second that appear, {\it prima facie},
to be far beyond the magnitude of fluctuations that could be expected from a `featureless' white-noise spectrum.
As discussed in the previous Section, the wavelet technique is very suitable for extracting such
structures, and has been applied in analogous analyses of time structures in photon emissions from
gamma-ray bursters~\cite{GRB}. The middle panel of Fig.~\ref{fig:w_img} shows the normalized wavelet power
spectrum, $|W_n(s)|^2/\sigma^2$, for the time series of the neutrino emission shown in the top panel.
The normalization by $1/\sigma^2$ gives a measure of the power relative to white
noise, and the colours represent the significance of the
feature compared to a white-noise spectrum. We see that the wavelet transform picks
out structures in the time series on time scales down to $\sim 2 \times 10^{-3}$~s.
Several of these structures in the time series have high significance, well
above the 95\% CL for a white-noise spectrum (indicated by red contours). These can be seen in 
the bottom panel of Fig.~\ref{fig:w_img}, where stuctures with time scales 
between
2~ms  and 3~ms are selected. We note, in 
particular, the series of 
structures appearing at times between 0.22 and 0.34~s after the start. 
There are also structures
in the band between 7~ms and 15~ms, and 
also appreciable power
at longer periods. Since we are interested in obtaining the best resolution 
possible, in
 the following 
 we focus 
 on the band corresponding to the smallest range of
the scales where significant power is seen, namely that  between  
2~ms and 3~ms~\footnote{However, we do note that these would be the most endangered
by directional averaging of neutrinos streaming in
all directions, as may occur in more complete 2D or 3D treatments of the neutrino transport.
In contrast, the time structures connected to the typical SASI and
convective timescales (tens of Hertz up to about 200 Hz) are likely
to survive even in 3D (though possibly with a somewhat reduced amplitude)~\cite{3D-1,3D-2}.
If the minimum period of neutrino variability were to increase, the sensitivity of our analysis
would decrease correspondingly.}.

Here, we investigate how these structures would be smeared out by the
energy-dependent refractive index
or by a stochastic spread in the velocities of different neutrinos with 
the same energy, the two possibilities
 described
 in the previous section.  Specifically,
we study the following possible energy dependences of the neutrino 
group velocity $v_{g\nu}$:
 \beq\label{LV1/2}
\frac{v_{g\nu}}{c}=1\pm\left(\frac{E}{M_{\nu 
{\rm LV}\frac{1}{2}}}\right)^{1/2} ,
\eeq
\beq\label{LV1}
\frac{v_{g\nu}}{c}=1\pm\frac{E}{M_{\nu {\rm LV}1}} ,
\eeq
\beq\label{LV2}
\frac{v_{g\nu}}{c}=1\pm\left(\frac{E}{M_{\nu {\rm LV}2}}\right)^2.
\eeq
%  {\bf Edward:} Here we add the $E^{1/2}$ Lorentz-invariance violation term in order to  trace a power-dependence of the phenomenon.  
We also investigate a possible stochastic effect which
may change the arrival times, $t$, of individual neutrinos,
assuming a Gaussian  probability distribution function:
\beq\label{stoch}
{\cal P}(t^{\rm stoch})=\frac{1}{\sigma\sqrt{2\pi}}\exp
 \left[-\frac{(t^{\rm stoch}-t)^2}
{2\sigma^2}\right],
\eeq
 as discussed 
above.
Here $\sigma =\gamma_{\, l} E^{\, l}$ with constants $\gamma_{\, l}$ and
$l=0,1,$ and $2$.

%We assume in our analysis that the neutrino data collected from a supernova
%explosion will consist of a list of individual neutrino events with
%measured energies $E_i$ and arrival times $t_i$. 
%The results of the simulation in~\cite{2D} are
%presented as a set of energy fluxes within time periods of durations $\simeq 3 - 5~{\rm ms}$, and each
%individual flux
%can be represented by a black-body spectrum with a given value of the mean energy. Knowing the
%mean and total energy of the flux in each time period, we assign 
%statistically to each neutrino in the simulation a specific time of emission and energy. 
%The distribution of one implementation of such neutrino time-energy assignments is presented in
%Fig.~\ref{fig:ET}. In order to estimate the sensitivities to novel effects in neutrino propagation as
%discussed below, we make a number of different implementations of the neutrino emission, all with
%independent
%statistical realizations of the thermal spectra. We then make wavelet transforms of these
%implementations,
%and analyze statistically the sensitivities to new physics that they may have.

\subsection{Sensitivity to an energy-dependent refractive index for neutrinos}

Lower limits on $M_{\nu {\rm LV}\frac{1}{2}}$, $M_{\nu {\rm LV}1}$, 
$M_{\nu {\rm LV}2}$ and $\gamma_l$ may be
calculated by requiring that the fine-scale time structures in the wavelet power spectrum 
do not disappear below the 95\% CL of significance for a signal above the white-noise 
power spectrum. Specifically, for
the models (\ref{LV1/2}),  (\ref{LV1}) and (\ref{LV2}) we apply to every neutrino event an 
energy-dependent time shift 
\beq\label{shiftTime1}
\Delta t=\tau_{\, l}E^{\, l},
\eeq
where 
\beq\label{deltaTau1}
\tau_{\,l}=\frac{L}{cM_{\nu {\rm LV}l}^l},
\eeq
and $l=\frac{1}{2},1,2$. We then vary $\tau_{\, l}$ ($M_{\nu {\rm LV}l}$) 
and follow the evolution of the
signal in the neutrino time series. If there is a non-trivial dispersive effect during propagation
from the source, it can be compensated by choosing the
``correct'' value of the time shift $\tau_{\, l}$, in which case the 
original time structure at the
source is recovered. On the other hand, dispersion at the source itself could not, in general,
be compensated by any choice of $\tau_{\, l}$. Quantitatively, the time 
structure of the supernova signal 
is recovered by maximizing the fraction of the scale-averaged power spectrum above the 95\% 
CL line. In order to calculate a lower limit on $\tau_{\, l}$ in any 
specific model,
we examine the fine-scale time structures that appear
above the 95\% CL in the bottom panel of Fig.~\ref{fig:w_img} and find
the value of the time-shift parameter (\ref{deltaTau1}) at which the signal above
the 95\% CL disappears. 

\begin{figure}[htb]
\begin{center}
\includegraphics[width=0.8\textwidth]{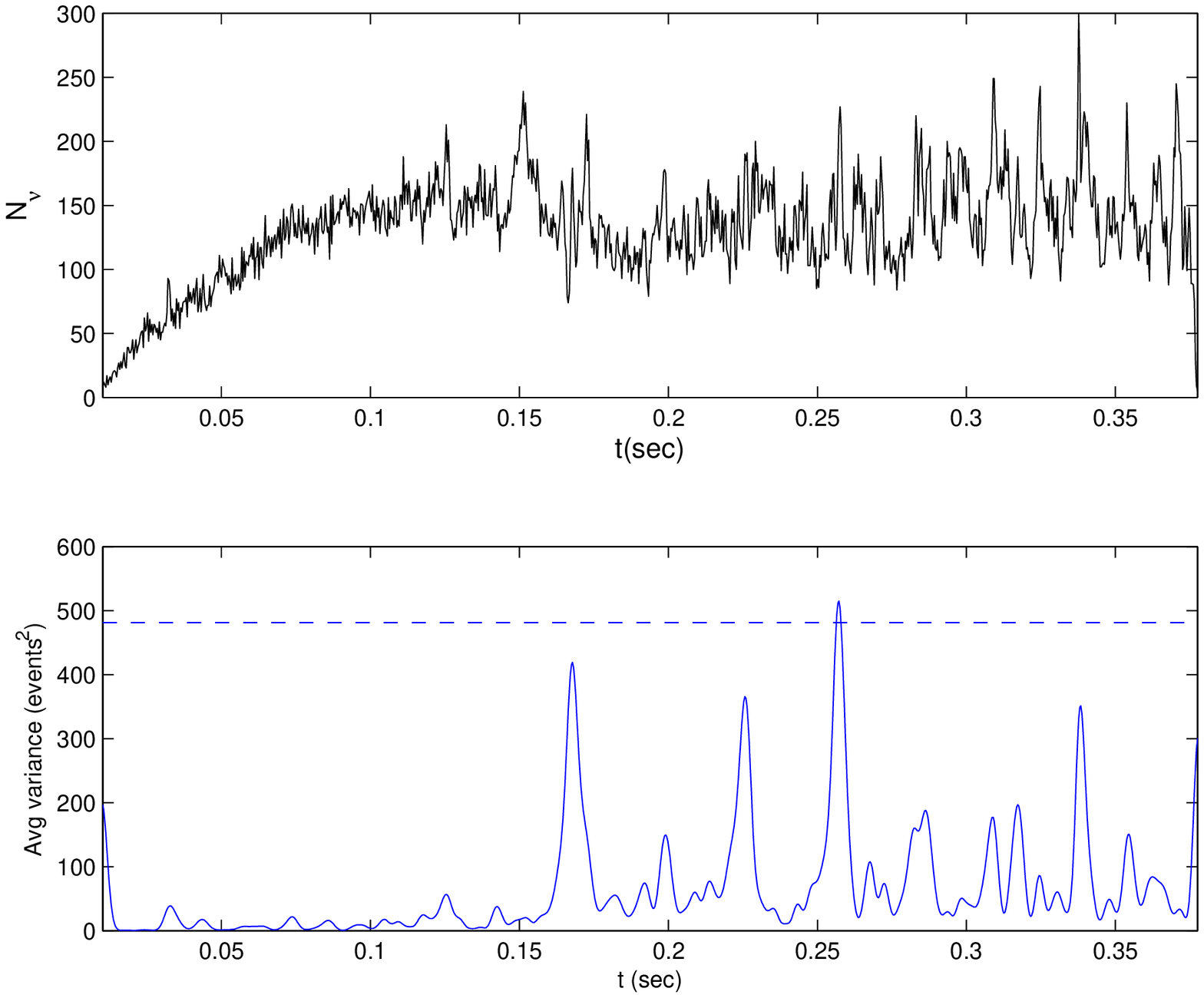}
\end{center}
\caption{\it The strength of the time-scale structure of the power spectrum averaged between
0.002 and 0.003~s disappears below the 95\% CL of significance after applying a linear 
energy-dependent time shift $\tau_{1} =4.2\times 10^{-5}$~(s/MeV).}
\label{fig:below}
\end{figure}

We first study a linearly energy-dependent neutrino refractive index
of the form $1 + (E/M_{\nu {\rm LV}1})$. 
Fig.~\ref{fig:below} displays the result of one simulation of the effect of such an
energy-dependent refractive index, sampled in 21 bins corresponding to different
time shifts $\tau_{1}$. The vertical axis shows the strengths of the emissions in the structures
with time scales between $2\times 10^{-3}$~s and $3 \times 10^{-3}$~s, applying a linear 
energy-dependent time shift $\tau_{1} =4.2\times 10^{-5}$~(s/MeV).
Looking at the structures that occur between 0.22 and 0.34~s after the start,
we find that the significant portions of these small-scale structures (those that
rise above the 95\% fluctuation level for white-noise background neutrino emission)
disappear for time
delays $\tau > 4.20 \times 10^{-5}$~(s/MeV), corresponding to $M_{\nu 
{\rm LV}1} > 2.45 \times
10^{13}$~GeV if a supernova distance $L$ of 10~kpc is assumed. 
This sensitivity is two orders of magnitude more sensitive than that found in~\cite{Harries}, namely
$M_{\nu {\rm LV}1} > 2.2 \times 10^{11}$~GeV, based on a one-dimensional 
simulation of a core-collapse
supernova that did not exhibit the small time-scale structures seen in Fig.~\ref{fig:w_img}.

\begin{figure}[htb]
\begin{center}
\includegraphics[width=0.8\textwidth]{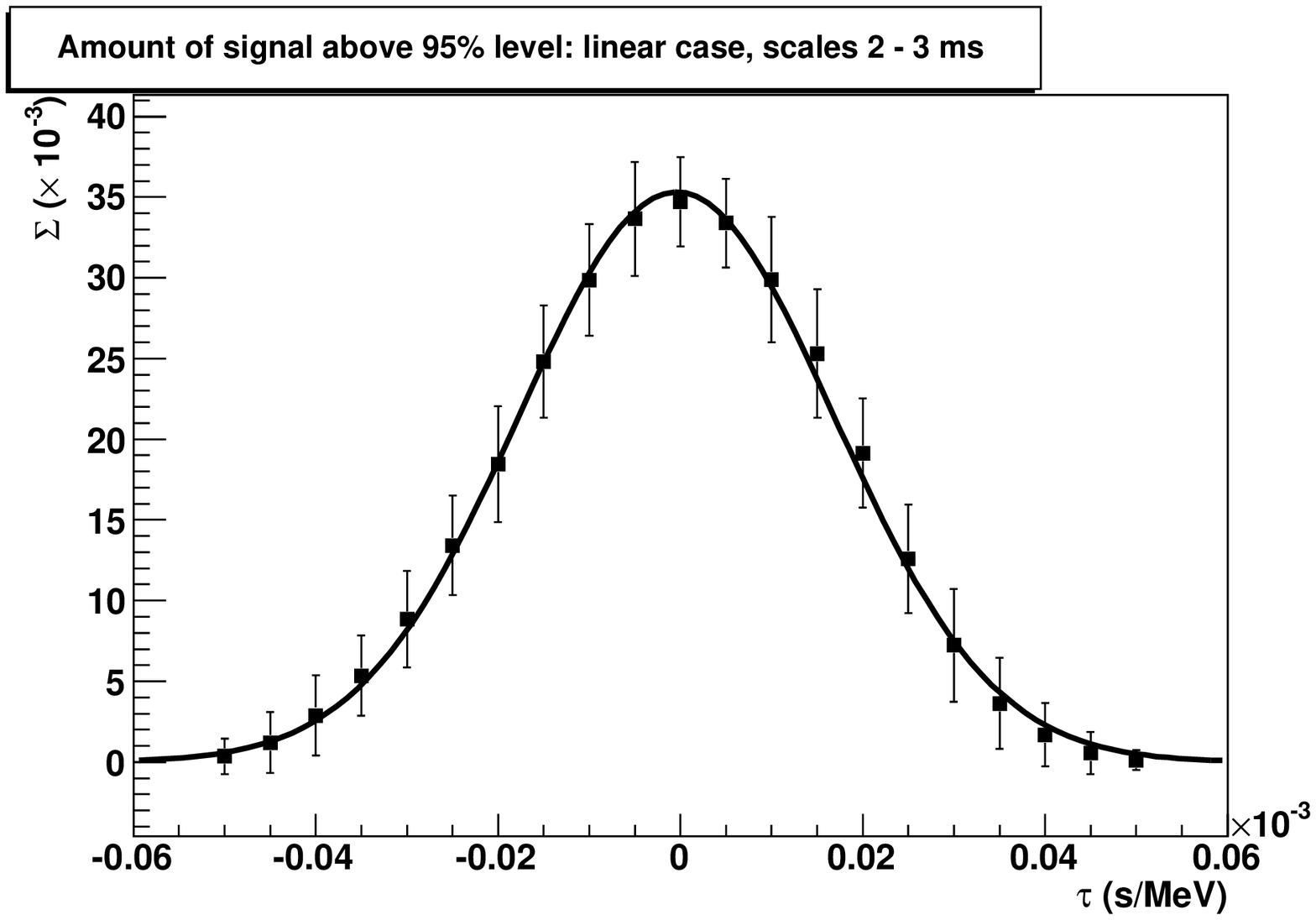}
\end{center}
\caption{\it A Gaussian fit to the amount $\Sigma$ of the short time-scale signal above the 95\%~CL,
calculated for 21 values of the shift parameters $\tau_1$. Each point is obtained as the average 
over 25 realizations of the time-energy assignments of individual neutrinos.}
\label{fig:E1Fit}
\end{figure}

We have repeated this exercise with 25 different statistical realizations of the
neutrino emission, calculating in each case the amount $\Sigma$ of the total signal above 
95\% CL for different values of $\tau_1$ sampled in 21 bins. The results of these 25 
realizations can be fit quite well by a Gaussian distribution, as seen in
Fig.~\ref{fig:E1Fit}. (In this and subsequent figures, we concentrate on the structures
with time scales between $2\times 10^{-3}$~s and $3 \times 10^{-3}$~s
that occur between 0.22 and 0.34~s after the start.) One can see that the 
position of the maximum, which defines the value of  $\tau$ that maximizes the time structures
in the signal and is expected to be zero, is indeed consistent with zero to within a precision 
of $10^{-6}$~(s/MeV), while the structures are washed out to below the $1\sigma$ level 
at
\beq\label{tau1limit}
\tau_{\, 1}=3.85 \, [3.95]\times
10^{-5}\ {\rm s/MeV} .
\eeq
where the number in square brackets $[ ... ]$ is obtained from a similar analysis of the superluminal case.
On the basis of this analysis, if significant time structures of the type found in the two-dimensional
simulation~\cite{2D} were to be seen in IceCube in neutrino data from a core-collapse
supernova at a distance of 10~kpc, one could conclude that
\begin{equation}
M_{\nu {\rm LV}1} \; > \; 2.68 \, [2.61] \times 10^{13}~{\rm GeV}.
\label{limit1}
\end{equation}
for a neutrino refractive index of the form $1 \pm (E/M_{\nu {\rm LV}1})$.
On the other hand, if no such structures were seen, inferring an upper 
limit on $M_{\nu {\rm LV}1}$ would 
require strong independent confirmation of the structures found in~\cite{2D}, in particular
by full three-dimensional simulations.

\begin{figure}[htb]
\begin{center}
\includegraphics[width=0.8\textwidth]{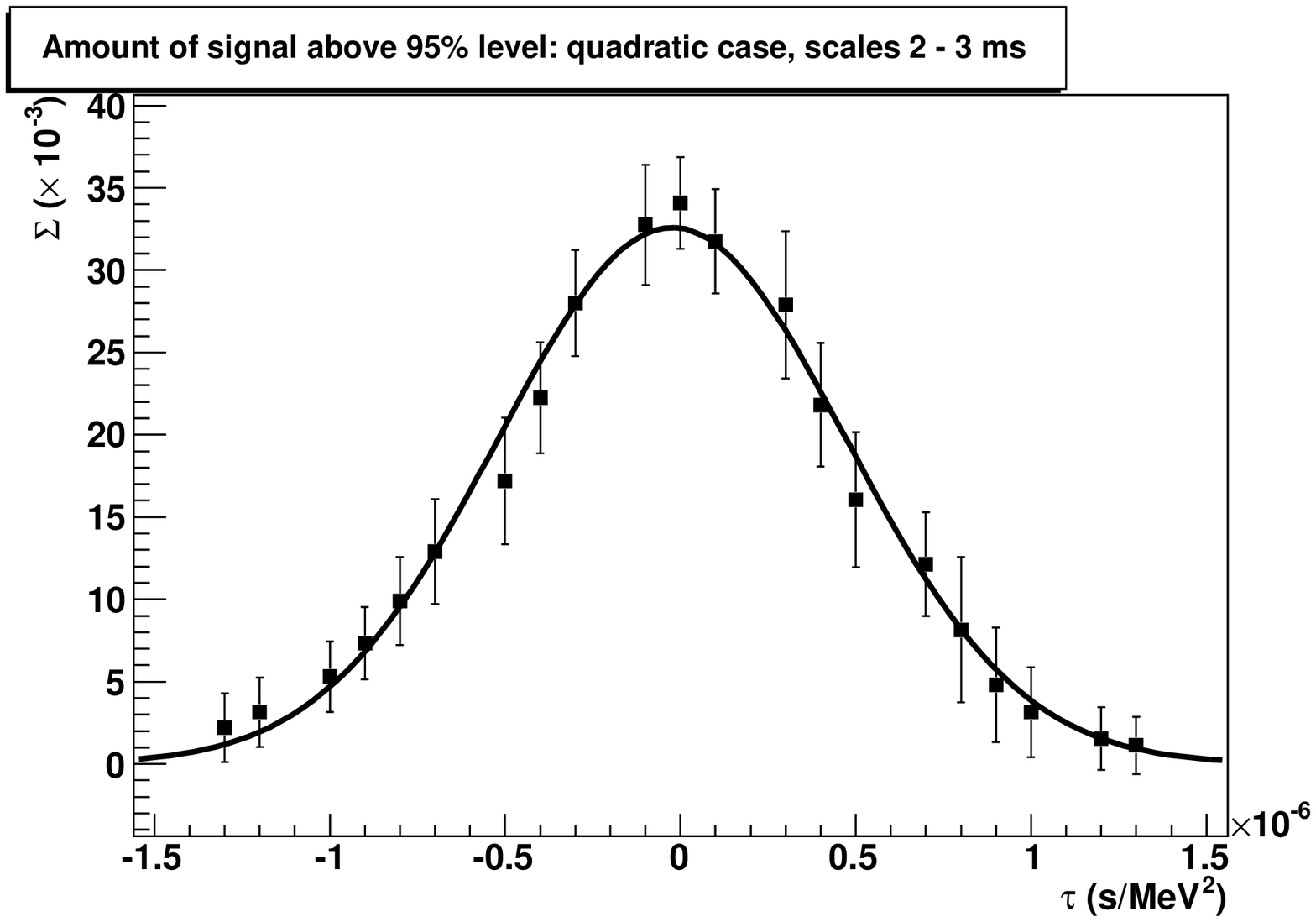}
\end{center}
\caption{\it A Gaussian fit to the amount $\Sigma$ of the short time-scale signal above the 95\%~CL
calculated for 21 values of the shift parameters $\tau_2$. Each point is obtained as the average 
over 25 realizations of the time-energy assignments of individual neutrinos.}
\label{fig:E2Fit}
\end{figure}

We have repeated this analysis for the case of a quadratic energy dependence in the refractive
index of the form $1 \pm (E/M_{\nu {\rm LV}2})^2$, finding the results shown 
in
Fig.~\ref{fig:E2Fit}. 
In this case, we find that the structures are washed out 
to below the $1\sigma$ level when
\beq\label{tau2limit}
\tau_2=1.10 \, [1.11] \times
10^{-6}\ {\rm s/MeV^2},
\eeq
where the number in square brackets is again obtained from a similar analysis of the superluminal case.
Hence, observation of significant time structures~\cite{2D} in IceCube would imply 
that
\begin{equation}
M_{\nu {\rm LV}2} \; > \; 0.97 \, [0.96] \times 10^{6}~{\rm GeV} ,
\label{limit2}
\end{equation}
if such structures were to be observed in a supernova explosion at 10~kpc, again with the proviso that
inferring an upper limit on $M_{\nu {\rm LV}2}$ would 
require strong confirmation of the structures found in~\cite{2D}, specifically by full three-dimensional
simulations.

\begin{figure}[htb]
\begin{center}
\includegraphics[width=0.8\textwidth]{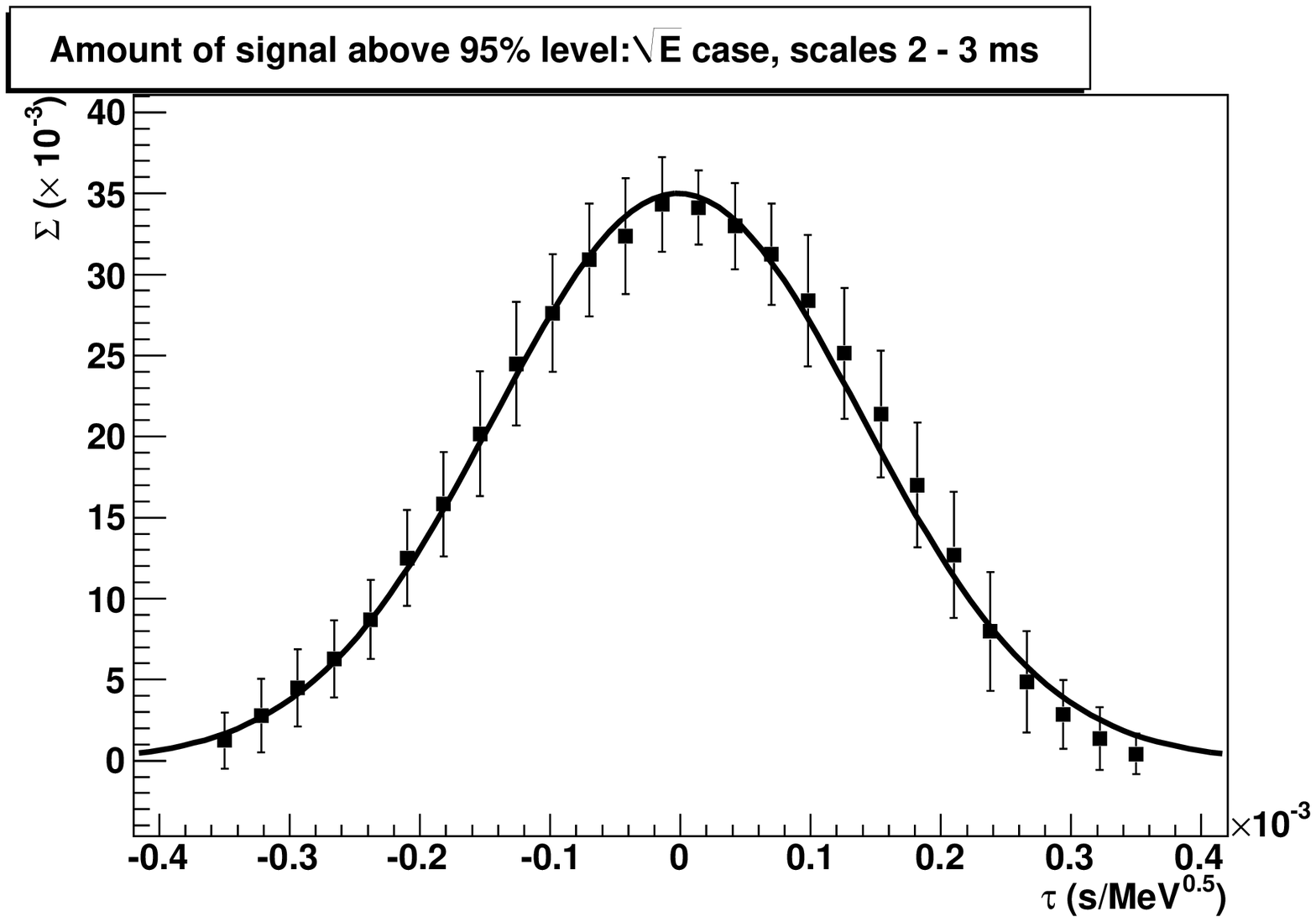}
\end{center}
\caption{\it A Gaussian fit to the amount $\Sigma$ of the short time-scale signal above the 95\%~CL
calculated for 26 values of the shift parameters $\tau_{\frac{1}{2}}$. Each point is obtained as the average 
over 25 realizations of the time-energy assignments of individual neutrinos.}
\label{fig:E05Fit}
\end{figure}

Finally, repeating this analysis for the case $1 \pm \sqrt{E/M_{\nu 
{\rm LV}\frac{1}{2}}}$, we find
the results shown in Fig.~\ref{fig:E05Fit}: 
\beq\label{tau1/2limit}
\tau_{1/2}=3.10 \, [3.15] \times
10^{-4}\ {\rm s/\sqrt{MeV}},
\eeq
(again, square brackets denote the
superluminal case) corresponding to a sensitivity to
\begin{equation}
M_{\nu {\rm LV}\frac{1}{2}} \; > \; 1.11 \, [1.07] \times 10^{22}~{\rm GeV}
\label{limithalf}
\end{equation}
if such structures were to be observed in a supernova explosion at 10~kpc.

\subsection{Sensitivity to a stochastic spread in neutrino velocities}

The possibility of a stochastic spread in the velocities of individual neutrinos with
the same energy can be investigated in a similar way. As discussed in Section~2,
the amount of velocity spread $\sigma$ might be energy-independent, and we consider this
possibility as well as possible linear and quadratic energy dependences of $\sigma$.
As in the case of an energy-dependent time delay, any of these possibilities would tend to
spread out and reduce the significances of the peaks found in the wavelet analysis shown in
Fig.~\ref{fig:w_img}. 

\begin{figure}[htb]
\begin{center}
\includegraphics[width=0.8\textwidth]{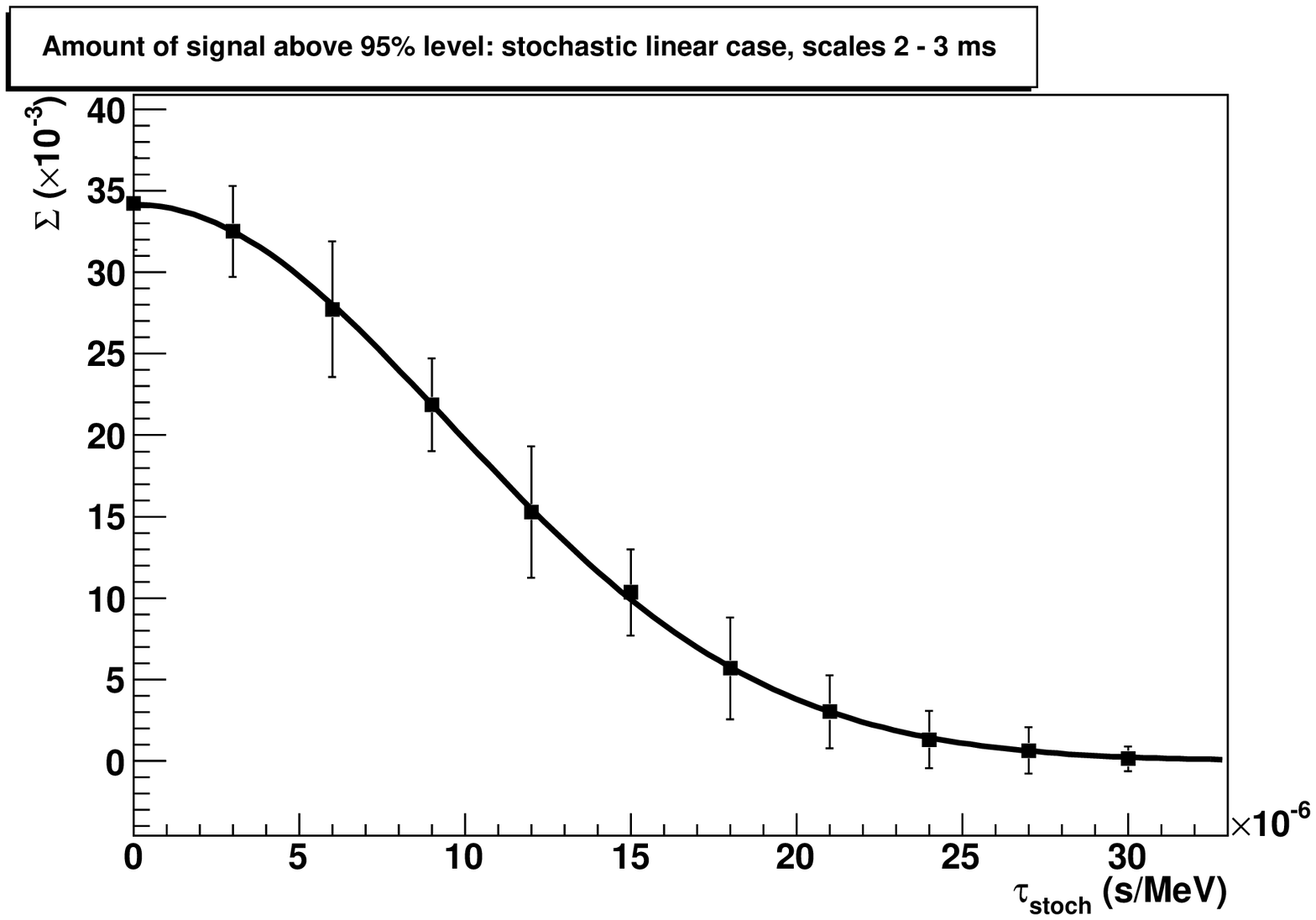}
\end{center}
\caption{\it A Gaussian fit to the amount $\Sigma$ of the short time-scale signal above the 95\%~CL
calculated for 11 values of the shift parameters $\tau_1^{\rm stoch}$. Each point is obtained as the average 
over 35 realizations of the time-energy assignments of individual neutrinos.}
\label{fig:stochE1Fit}
\end{figure}

\begin{figure}[htb]
\begin{center}
\includegraphics[width=0.8\textwidth]{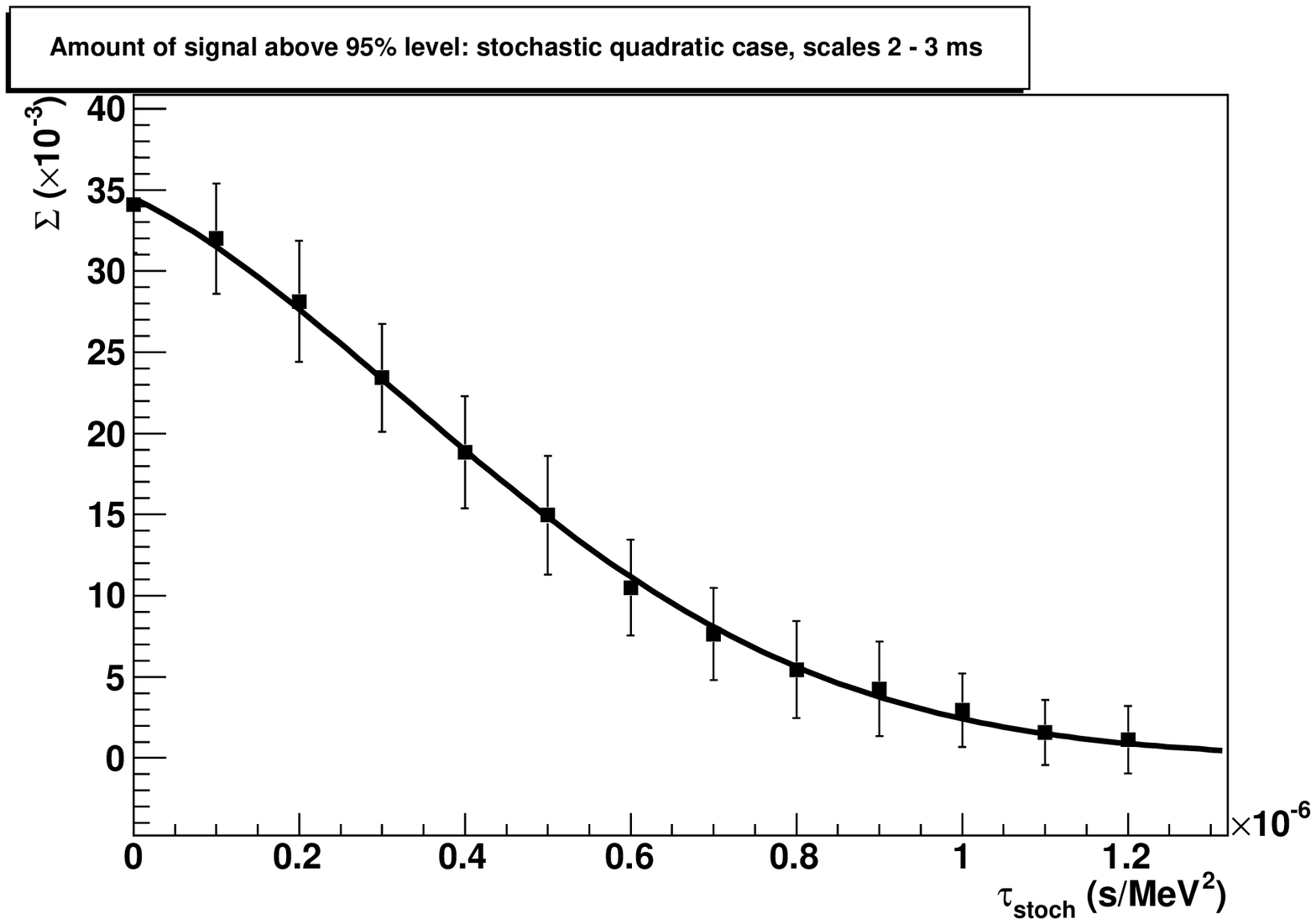}
\end{center}
\caption{\it A Gaussian fit to the amount $\Sigma$ of the short time-scale signal above the 95\%~CL
calculated for 13 values of the shift parameters $\tau_2^{\rm stoch}$. Each point is obtained as the average 
over 35 realizations of the time-energy assignments of individual neutrinos.}
\label{fig:stochE2Fit}
\end{figure}

This effect is seen for $\sigma \propto E$ in 
Fig.~\ref{fig:stochE1Fit}, and for $\sigma \propto E^2$ in
Fig.~\ref{fig:stochE2Fit}. In each case, we plot
results obtained from 35 independent statistical simulations of the neutrino emission signal.
We see that the wavelet
peaks are reduced below the 95\% CL white-noise level for 
\beq\label{tauStoch1limit}
\tau_{\rm stoch1}=2.16\times
10^{-5}\ {\rm s/MeV}
\eeq
in the linear case, 
\beq\label{tauStoch2limit}
\tau_{\rm stoch2}=9.56\times
10^{-7}\ {\rm s/MeV^2}
\eeq
in the quadratic case, and
\beq\label{tauStoch0limit}
\tau_{\rm stoch0}=3.59\times
10^{-4}\ {\rm s}
\eeq
in the energy-independent case~\footnote{For comparison, we note that the OPERA result~\cite{OPERA}
also provides an upper limit on the stochastic spread of about 10~ns, though after propagation over a
much shorter distance $\sim 730$~km. However, using equations (\ref{timegauss}) to (\ref{alphavalue})
we find that neither OPERA nor SN1987a gives competitive constraints on 
the scales $M_{\nu {\rm LV}l}$.}.
These sensitivities correspond to $M_{\nu \widetilde{{\rm LV}1}} > 4.78 
\times 10^{13}$~GeV and $M_{\nu \widetilde{{\rm LV}2}} >
1.04 \times 10^{6}$~GeV in the energy-dependent cases, for a supernova explosion at 10~kpc.

\begin{figure}[htb]
\begin{center}
\includegraphics[width=0.8\textwidth]{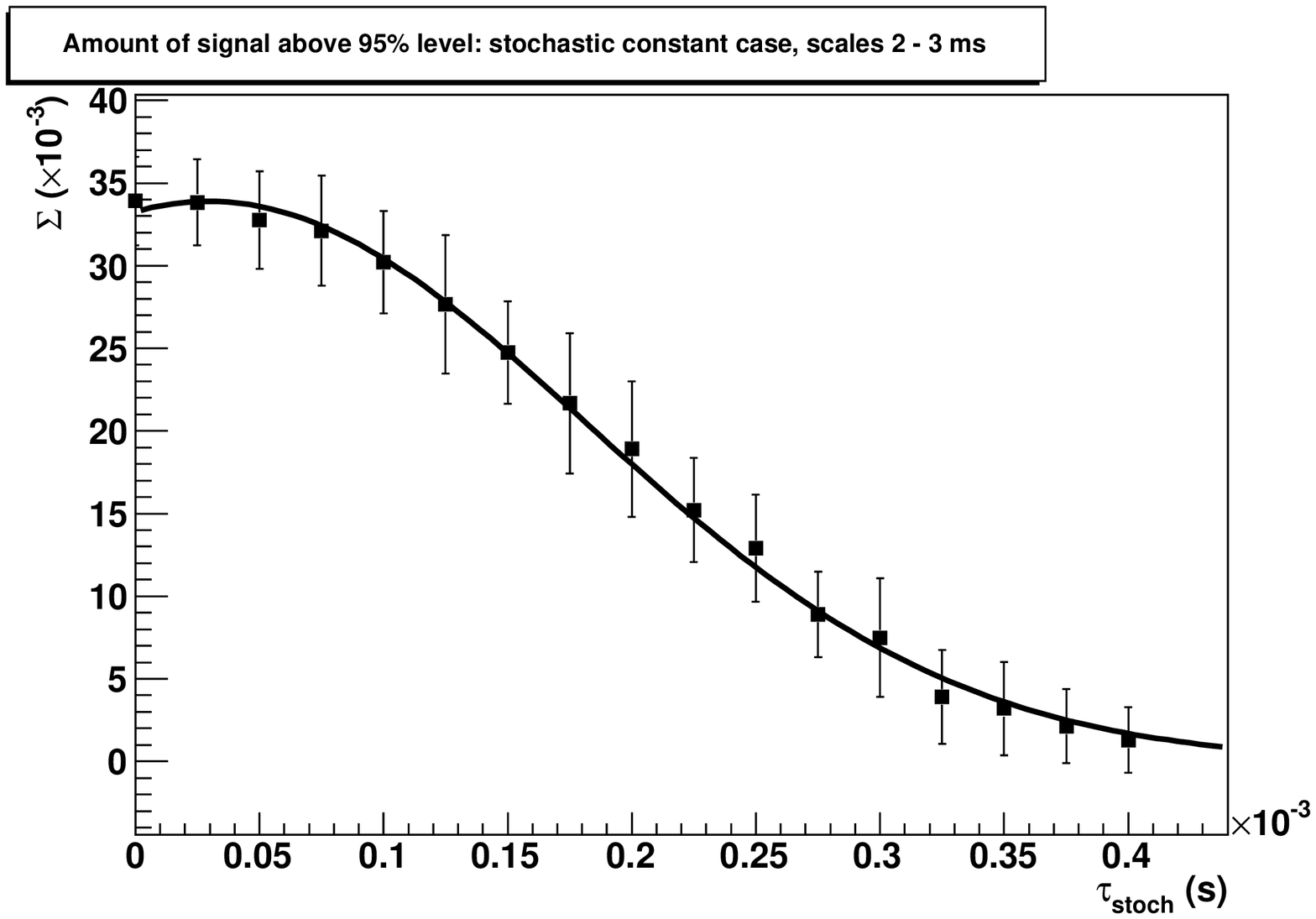}
\end{center}
\caption{\it A Gaussian fit to the amount $\Sigma$ of the short time-scale signal above the 95\%~CL
calculated for 17 values of the shift parameters $\tau_0^{\rm stoch}$. Each point is obtained as the average 
over 35 realizations of the time-energy assignments of individual neutrinos.}
\label{fig:stochE0Fit}
\end{figure}

\section{Conclusions and Prospects}

We have shown that the existence of structures with short time scales in the neutrino
emission from a core-collapse supernova, as suggested by two-dimensional
simulations~\cite{2D}, would open up new possibilities for probing
aspects of the propagation of neutrinos that lie far beyond the reach of terrestrial experiments,
and up to two orders of magnitude beyond the sensitivity provided by previous analyses 
based on one-dimensional supernova simulations. This increased sensitivity holds
for possible square-root, linear and quadratic dependences of the neutrino refractive index, and
for both energy-independent and linear or quadratically-dependent stochastic spreads
in the velocities of different neutrinos with the same energy. 

Specifically, if such short time structures are seen in a supernova explosion at a distance of
10~kpc, one could infer that
\begin{eqnarray}
M_{\nu {\rm LV}\frac{1}{2}}  & > & 1.11 \, [1.07] \times 10^{22}~{\rm GeV} , \\
M_{\nu {\rm LV}1} & > & 2.68 \, [2.61] \times 10^{13}~{\rm GeV} , \\
M_{\nu {\rm LV}2} & > & 0.97 \, [0.96] \times 10^{6}~{\rm GeV} , \\
M_{\nu \widetilde{{\rm LV}1}} & > & 4.78 \times 10^{13}~{\rm GeV} , \\
M_{\nu \widetilde{{\rm LV}2}} & > & 1.04 \times 10^{6}~{\rm GeV} ,
\label{results}
\end{eqnarray}
where the numbers in square brackets correspond to the superluminal case, and
the last two limits correspond to the possible effects of stochastic fluctuations. In 
the case of an energy-independent stochastic spread, one could infer that
$\tau_{\rm stoch0} < 3.59 \times 10^{-4}$~s.

If such short time structures are not seen, many checks would be necessary before one
could conceivably claim observation of any unconventional effect in neutrino
propagation. In particular, it would be necessary to validate the predictions of the two-dimensional
core-collapse supernova simulation on which this analysis is based, specifically by
confirming that short time structures are also found in full three-dimensional
simulations~\cite{3D-1,3D-2}. We hope that the interesting sensitivity to new neutrino physics
discussed in this paper - not to mention OPERA~\cite{OPERA} - will add to the motivation to develop further such simulations
and derive robust predictions for neutrino emissions from core-collapse supernovae. 

\section*{Acknowledgements}

The work of J.E. and N.E.M. was supported partly by the London Centre for
Terauniverse Studies (LCTS), using funding from the European Research
Council via the Advanced Investigator Grant 267352.
H.-T.J. acknowledges support by the Deutsche Forschungsgemeinschaft
through the Transregional Collaborative Research Centers SFB/TR~27
``Neutrinos and Beyond'' and SFB/TR~7 ``Gravitational Wave Astronomy'',
and the Cluster of Excellence EXC~153 ``Origin and Structure of the Universe''
({\tt http://www.universe-cluster.de}). The supernova simulations were 
possible by computer time grants at the John von Neumann Institute for
Computing (NIC) in J\"ulich, the H\"ochst\-leistungs\-re\-chen\-zentrum
of the Stuttgart University (HLRS) under grant number SuperN/12758,
the Leib\-niz-Re\-chen\-zentrum M\"unchen, and the RZG in Garching.
We are grateful to Andreas Marek for providing the neutrino data.

%\newpage

\end{document}